\documentstyle[twoside,fleqn,espcrc2,epsfig]{article}

\newcommand{\chup}{$\chi U\phi$}

\newcommand{\eqref}[1]{(\ref{#1})}

\newcommand\fdfig[1]{%
  \psfig{file=#1,angle=90,width=\hsize,bbllx=65,bblly=15,bburx=540,bbury=780}}%

\newcommand{\AmS}{{\protect\the\textfont2
  A\kern-.1667em\lower.5ex\hbox{M}\kern-.125emS}}

\title{New universality class of chiral symmetry breaking
       in the strongly coupled U(1) $\chi U \phi$ model}

\author{W.~Franzki and J.~Jers{\'a}k
\address{Institut f{\"u}r
Theoretische Physik E, RWTH Aachen, Germany}}
       
\begin{document}

\begin{abstract}
We describe a 4D U(1) lattice gauge theory with charged scalar $\phi$ and
fermion $\chi$ matter fields ($\chi U \phi$ model). At sufficiently strong
gauge coupling, the chiral symmetry is broken and the mass of the unconfined
composite fermion $F = \overline{\chi} \phi $ is generated dynamically by
gauge interaction. The scalar supresses this symmetry breaking and induces a
line of second order transitions with scaling properties similar to the
Nambu--Jona-Lasinio model. However, in the vicinity of a particular,
tricritical point the scaling properties are different. Here we study the
effective Yukawa coupling between the massive fermion and the Goldstone boson.
The perturbative triviality bound of Yukawa models is nearly saturated. The
theory is similar to strongly coupled Yukawa models except the occurrence of
an additional state -- a gauge ball of mass $m_S \simeq 1/2 m_F$. This, and
non-classical values of tricritical exponents suggest that at the tricritical
point the $\chi U \phi$ model constitutes a new universality class.
Nevertheless, it might be a microscopic model for the Higgs-Yukawa mechanism
of symmetry breaking.
\end{abstract}

\maketitle

\section{INTRODUCTION}

Common mechanisms of fermion mass generation in QFT belong to one of the two
generic types: First, the chiral symmetry is broken by the scalar field and
the fermion mass is a consequence of the Yukawa coupling, like in the
Higgs-Yukawa sector of the standard model. The four-fermion theory belongs to
this type through universality, though the scalar field is auxiliary in
this case. Second, the chiral symmetry is broken by a strong gauge interaction
accompanied by confinement of the fermions acquiring mass, like in QCD or
technicolor, or by a massless photon, like in the noncompact QED. No scalar
field is involved.

A new generic mechanism, different from the above ones, has been suggested in
Ref.~\cite{FrJe95a}. The exemplary ``\chup\ model'' consists of a charged
fermion field $\chi$ with strong vectorlike coupling to compact U(1) gauge
field $U$, and a scalar field $\phi$ of the same charge.

In the \chup\ model, the scalar field $\phi$ helps to solve two problems.
First, it shields the fermion charge and gives rise to an unconfined, i.e.
physical massive fermion $F = \phi^\dagger\chi$ in the phase with chiral
symmetry broken dynamically by the gauge interaction (Nambu phase).  Second,
the scalar {\em suppresses} this symmetry breaking and at sufficiently strong
gauge coupling induces a second order transition to a chiral symmetric
phase, thus opening a way to continuum. There is a particular point of the
corresponding phase transition, a tricritical point, at which the model
defines a new universality class.

We summarize the results of our extensive systematic investigations of this
model in four dimensions by means of numerical simulations with dynamical
fermions.  The detailed account is given in \cite{FrJe98a,FrJe98b}. Here we
discuss the particular properties of the new universality class governing the
tricritical point.


\section{THE $\chi U \phi$ MODEL}

The model is defined by the action
\begin{eqnarray*}
  S &=& S_\chi + S_U + S_\phi \; , \\ 
  S_\chi & = & \frac{1}{2} \sum_x
   \overline{\chi}_x \sum_{\mu=1}^4 \eta_{x\mu} (U_{x,\mu} \chi_{x+\mu} -
   U^\dagger_{x-\mu,\mu} \chi_{x-\mu}) \\ 
         & &+{am_0} \sum_x \overline{\chi}_x\chi_x\; ,\\ 
  S_U & = & -\beta \sum_P \cos(\Theta_P)\; , \\ 
  S_\phi & = & - {\kappa}
   \sum_x \sum_{\mu=1}^4 (\phi^\dagger_x U_{x,\mu} \phi_{x+\mu} + h.c.).
\end{eqnarray*}
Here $\Theta_P \in [0,2\pi)$ is the plaquette angle. We use staggered
fermions. The complex scalar field is constrained, $|\phi| = 1$. As
its ``hopping parameter'' $\kappa$ increases, it drives the model into
the usual Higgs phase.

\begin{figure}
  \begin{center}
    \fdfig{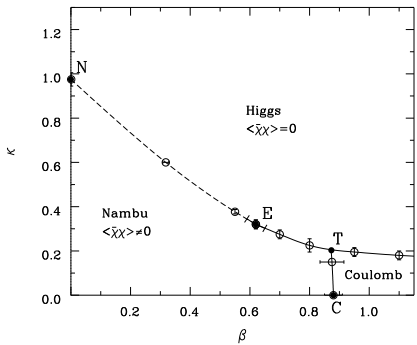}%
    \caption[xxx]{%
      Phase diagram of the \chup\ model in the chiral limit, $m_0 =
      0$. The $\beta = 0$ limit corresponds to the NJL model. The NE
      line is a line of second order phase transitions. Other lines
      are lines of first order transitions. In the Nambu phase, chiral
      symmetry is broken dynamically and the fermion $F =
      \phi^\dagger\chi$ is massive.}
    \label{fig:pd4d2}
  \end{center}
\end{figure}

We note that there is no Yukawa coupling between $\chi$ and $\phi$, as both
fields have the same charge. The phase diagram at $m_0= 0$ is shown in
Fig.~\ref{fig:pd4d2}. Numerical simulations have to be carried out at
nonvanishing $m_0$ and an extrapolation to the chiral limit $m_0 = 0$
performed.

\section{TRICRITICAL POINT E}
Point E is far away from any limit case and it does not appear to be
accessible by any reliable analytic method, neither on the lattice nor in
continuum. It is ``tricritical'' because in the full parameter space
(including $am_0$) there are, apart from NE, two further second order ``wing''
lines entering E from the positive and negative $am_0$ directions. The
existence of a common point E of these three second order lines is neither
predicted nor understood. The evidence is purely numerical, but quite strong
\cite{FrJe98b}. Its position is $\beta_{\rm E} = 0.62(3), \kappa_{\rm E} =
0.32(2)$.

The importance of the point E roots in the experience from statistical
mechanics that a tricritical point belongs to a universality class different
from that of any of the second order lines entering into it. The whole NE line
except the point E corresponds to the same continuum model as the point N, the
NJL model.  The gauge field is presumably auxiliary and the model is therefore
of limited interest there. However, the point E is expected to be different,
gauge field playing an important role.

To verify this expectation, we have investigated critical exponents and
spectrum of the model in the vicinity of E.  Here we only mention the found
value $\nu_t \simeq 1/3$ of the correlation length tricritical exponent. It
differs from the prediction of the classical theory of tricritical points,
indicating important role of quantum fluctuations at the point E
\cite{FrJe98b}.

\section{SPECTRUM AT E AND ITS \hfill \\ INTERPRETATION}
Some insight into the physics of the continuum limit at the point E is
provided by the spectrum and its scaling behavior. The massive physical
fermion $F$, as well as other physical states, are composite.  The interaction
between them is due to the van der Waals remnant of the fundamental
interactions. We shall attempt a possibly somewhat naive but illustrative
interpretation of the found states in terms of the fundamental fields.

The fermion mass $am_F > 0$ decreases when the NE line is approached from the
Nambu phase. The data are consistent with expected vanishing of $am_{\rm F}$
in the Higgs phase in the infinite volume limit. One can consider a simple
composite picture of the fermion F and of its mass:
\begin{equation}
                     m_F \simeq \mu_\phi + m_\chi - E_B.
   \label{F-anatomy}
\end{equation}
Here $\mu_\phi$ is the mass of the scalar $\phi$ diverging at $\kappa = 0$.
The constituent mass of the $\chi$-fermion, $m_\chi$, is nonzero in the Nambu
phase even in the chiral limit because of the chiral symmetry breaking. $E_B$
is the binding energy, due to the gauge interaction between both constituents.
When at fixed $\beta$ in the Nambu phase the phase transition is approached by
increasing $\kappa$, two things happen in (\ref{F-anatomy}): as $\phi$ gets
lighter, $\mu_\phi$ decreases and the chiral symmetry breaking is suppressed.
Therefore also $m_\chi$ decreases. At critical $\kappa$ the mass $m_F$ in
(\ref{F-anatomy}) vanishes. Above the critical $\kappa$, the mass $m_F$ is
zero like in the standard model with vanishing Yukawa couplings.

Further we find in the Nambu phase several $\overline{\chi}$, $\chi$ bound
states. One of them is the obligatory pseudoscalar Goldstone boson $\pi$ with
the dependence on $am_0$ as required by current algebra.

Most interesting is the neutral scalar ($S$-boson). Its mass $am_S$ vanishes
on the wing critical lines, i.e. at the endpoints of the Higgs phase
transition at small $\beta$. It is seen appearing as a composite of
$\phi^\dagger$ and $\phi$ in the correlation function of the operator $ O_1 =
\phi^\dagger_x U_{x,\mu} \phi_{x+\mu}|_{scalar}$.

In the Higgs phase, at large $\beta$, $S$-boson can be identified with the
Higgs boson as obtained in the unitary gauge. Below the confinement-Higgs
phase transition, in particular in the Nambu phase, the interpretation of the
$S$-boson is different. As the unitary gauge is no more applicable, one
possibility is to see it as a $\phi^\dagger \phi$ bound state, i.e. as a
``meson'' consisting of two confined scalar ``quarks''. 

But it has been observed recently \cite{FrJe98b} that below the
confinement-Higgs phase transition the $S$-boson shows up also in the
gauge-ball correlation function $ O_2 = \cos \Theta_{P} |_{scalar}$, as well
as in the mixed correlation between $O_1$ and $O_2$. The amplitudes are as
large as in the correlation function of $O_1$. On the other hand, a
contribution of the $S$-boson to still another scalar channel, the
$\overline{\chi}\chi$ one, has not been detected and is thus very small. In
fact, no mass (``$\sigma$-meson'') has been detected in that channel.  Then a
scalar gauge ball is expected to be the lightest scalar in the Nambu phase,
because the $\phi^\dagger \phi$ bound state must get heavy when $\mu_\phi$
grows with decreasing $\kappa$. This interpretation is supported by the fact
that a light scalar is seen in the correlation function $O_2$ along the whole
phase boundary ETC.  Along TC it is clearly a gauge ball.

Therefore we conclude that the new mechanism of chiral symmetry breaking we
describe is accompanied by the presence of a scalar gauge ball in the
spectrum.

\section{MICROSCOPIC MODEL OF \hfill \\ A YUKAWA THEORY}
In the vicinity of the phase transition line NE we have determined the
renormalized Yukawa coupling $y_{\rm R}$ between $F$ and $\pi$. It is obtained
from the three-point function of the corresponding composite operators and
thus can be interpreted as an effective Yukawa coupling, which would describe
the interaction between $F$ and $\pi$ in the regime where their composite
structure can be neglected. We have found that at a fixed fermion mass $am_F$
the Yukawa coupling increases when $\beta$ decreases and the point E is
approached.

In the interval $y_{\rm R} \simeq 2 - 5 $, where the Yukawa coupling is
determined reliably, we have compared our results with the curve of maximal
renormalized coupling, resulting from the first order perturbative calculation
in the Yukawa model (triviality bound).  It turns out that, close to E, the
data are only slightly below such a curve. Therefore close to the tricritical
point the \chup\ model can be used as a microscopic model of an effective
strongly coupled Yukawa theory \cite{FrJe98a}.

We speculate that the suggested mechanism might be of some use for an
explanation of the large top quark mass beyond the standard model.

\section{ACKNOWLEDGEMENTS} 
The computations have been performed at the RWTH Aachen and HLRZ J\"ulich. We
thank HLRZ J\"ulich for hospitality.  The work was supported by DFG.


\bibliographystyle{wunsnot}   


\end{document}